# Phonon-Enhanced Nonlinearities in Hexagonal Boron Nitride

**Jared S. Ginsberg[1,#,*], M. Mehdi Jadidi[1,#], Jin Zhang[2,#,*], Cecilia Y. Chen[3], Nicolas Tancogne-Dejean[2], Sang Hoon Chae[4,5,6], Gauri N. Patwardhan[1,7], Lede Xian[2], Kenji Watanabe[8], Takashi Taniguchi[9], James Hone[4], Angel Rubio[2,10,*], and Alexander L. Gaeta[1,3,*]**

[1]*Department of Applied Physics and Applied Mathematics, Columbia University, New York, New York 10027, USA*
[2]*Max Planck Institute for Structure and Dynamics of Matter and Center for Free-Electron Laser Science, Hamburg 22761 Germany*
[3]*Department of Electrical Engineering, Columbia University, New York, New York 10027, USA*
[4]*Department of Mechanical Engineering, Columbia University, New York, New York 10027, USA*
[5]*School of Electrical and Electronic Engineering, Nanyang Technological University, Singapore 639798, Singapore*
[6]*School of Materials Science and Engineering, Nanyang Technological University, Singapore 639798, Singapore*
[7]*School of Applied and Engineering Physics, Cornell University, Ithaca, New York 14853, USA*
[8]*Research Center for Functional Materials, National Institute for Materials Science, 1-1 Namiki, Tsukuba 305-0044, Japan*
[9]*International Center for Materials Nanoarchitectonics, National Institute for Materials Science, 1-1 Namiki, Tsukuba 305-0044, Japan*
[10]*Center for Computational Quantum Physics, Simons Foundation Flatiron Institute, New York, NY 10010 USA*

[#]These authors contributed equally to this work
* *jsg2208@columbia.edu (J.S.G.), jin.zhang@mpsd.mpg.de (J.Z.), angel.rubio@mpsd.mpg.de (A.R.), and alg2207@columbia.edu (A.L.G.)*

**Abstract:** We investigate optical nonlinearities that are induced and enhanced due to the strong phonon resonance in hexagonal boron nitride. We predict and observe large sub-picosecond duration signals due to four-wave mixing (FWM) during resonant excitation. The resulting FWM signal allows for time-resolved observation of the crystal motion. In addition, we observe enhancements of third-harmonic generation with resonant pumping at the hBN transverse optical phonon. Phonon-induced nonlinear enhancements are also predicted to yield large increases in high-harmonic efficiencies.

## Introduction

Parametric optical processes in solids can provide a window into the optical susceptibility, band-structure, and underlying symmetries of crystals, each of which can dramatically affect the nonlinear frequency-conversion process[1–3]. Symmetries, more so than any other factor, dictate the allowed higher-order processes in a given nonlinear system[4]. These properties become frequency independent far from any resonances, as is the case in the visible and near-infrared regime where many high-order harmonic generation measurements take place[5]. However, in the mid-infrared regime, polar crystals support lattice collective oscillations that can be resonantly driven by an optical field. At frequencies near these phonon resonances the linear optical response of the crystal is significantly modified, manifesting for example as a peak in the real permittivity[6,7]. These ionic modes can alter the symmetry properties of the crystal, leading to transient nonlinear optical effects such as those observed in $SrTiO_3$, which can be driven into a metastable non-centrosymmetric state following prolonged exposure to a phonon-resonant pump[8]. Under increased resonant excitation using femtosecond laser pulses, the amplitude of the ionic motion can become nonlinear with the incident field strength. For bulk materials such as $LiNbO_3$ and GaAs, phonon-induced enhancements of optical nonlinearities[9–12] occur in this regime.

A strong phonon resonance in the mid-IR is present in the van der Waals crystal hexagonal boron nitride (hBN), with a transverse optical (TO) phonon mode at 7.3 μm free-space wavelength (170 meV)[13]. The relatively light constituent atoms of hBN make this one of the most energetic TO phonons, accessible by ultrafast table-top lasers. hBN has an energetically favorable AA' stacked lattice in equilibrium, with alternating boron and nitrogen atoms sitting one on top of the other. An illustration of the resonantly driven, in-plane displacement of atoms for the TO ($E_{1u}$) mode of hBN[14] is presented in Figure 1a. At the point where the photon and phonon dispersion curves meet, an anti-crossing emerges in the hBN band structure, and the crystal hosts new hybrid modes called phonon-polaritons[15]. These have been the subject of intense study due to their long-range propagation[16].

Using time-resolved measurements, we confirm that when TO phonons of hBN undergo oscillations as indicated by transient four-wave mixing (FWM) signals near the second harmonic generation (SHG) wavelength, which is forbidden with a single beam in a bulk sample at equilibrium[17–19]. The FWM signal is studied as a function of the power and polarization of both the phonon-inducing pump and harmonic-generating probe, from which preferential symmetry axes are identified. Moreover, the natural hyperbolicity of the hBN TO phonon makes it an attractive platform for tight confinement of optical energy, and therefore for enhancing nonlinearities and light-matter interactions within relatively large volumes[6]. We extend the scope of these light-matter interactions to a higher order in mid-IR power by exploiting the strong hyperbolic confinement for even greater electron-phonon coupling. Specifically, in this work we show enhanced emission from the phonon-electron contributions to optical third-harmonic generation (THG) in hBN. We theoretically predict and demonstrate experimentally the nonlinear response of thin hBN crystals associated with this TO phonon mode at 7.3 μm. By sweeping a significant bandwidth of the mid-IR we demonstrate a greatly enhanced on-resonance phononic contribution to THG when hBN is pumped at its TO phonon.

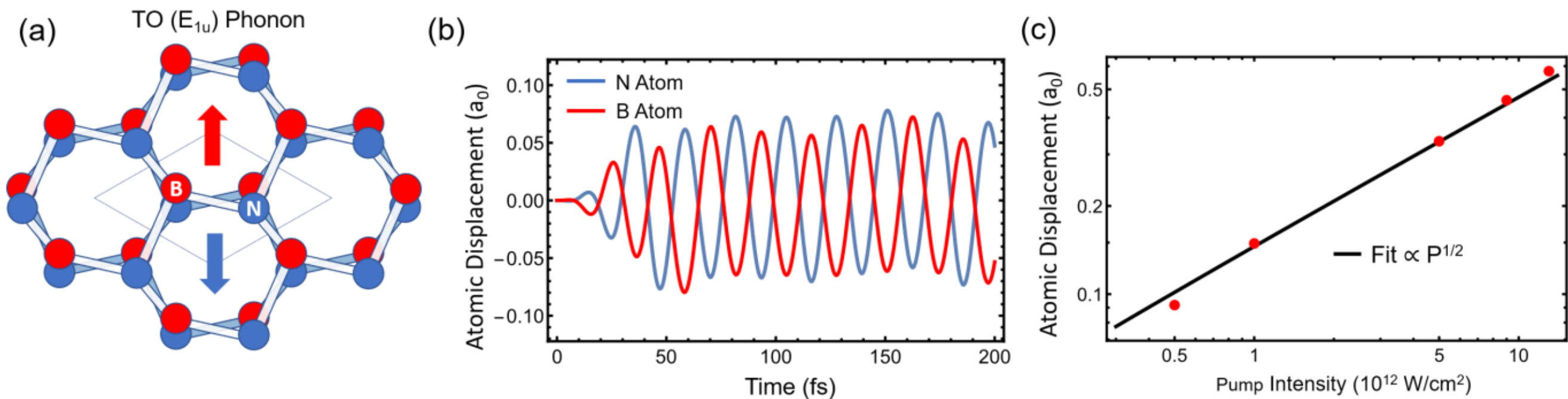

Figure 1: Atomic motion and atomic displacement associated with resonant driving of the TO ($E_{1u}$) phonon mode. (a) Honeycomb lattice arrangement of hexagonal boron nitride. Arrows illustrate the motion of atoms under resonant optical excitation. The two species move oppositely from each other in plane and across all layers for the IR-active TO ($E_{1u}$) mode. (b) Simulated atomic displacements of boron and nitrogen ions in TO ($E_{1u}$)-excited hBN. A 25-fs FWHM, 1 x $10^{12}$ W/cm$^2$ pulse excites the lattice dynamics. The TDDFT simulations do not include any damping terms through which to estimate the relaxation time. (c) Peak amplitude of atomic displacements as a function of pump intensity, fit to $I^{1/2}$ with a small linear-in-intensity correction. Displacements nearing 10% of the equilibrium lattice constant are achievable before the onset of damage.

# Results

**Phonon-Enhanced Four-Wave Mixing:** We first characterize theoretically the ionic displacements in bulk hBN under resonant excitation with 25-fs FWHM pulses by performing time dependent density-functional theory (TDDFT) simulated atomic oscillations spanning 200-fs, or roughly 8 times the theoretical pulse duration (see Figure 1b). For a modest input intensity of 1.5 x $10^{11}$ W/cm$^2$, we estimate that the phonon amplitude is 1% of the equilibrium lattice constant. While the period of the lattice oscillation is 25-fs, which is consistent with the expected phonon frequency, the relaxation time cannot be theoretically determined due to a lack of dissipative pathways. The amplitudes of atomic motion are plotted as a function of pump intensity in Figure 1c. The displacements predicted by TDDFT calculations are fit by $I^{1/2}$ with deviations appearing at large intensities and reach nearly 10% of the equilibrium lattice constant (2.5 Å) [20] at 10 TW/cm$^2$. The time-dependent electronic current is extracted, and from this we generate the theoretical harmonic spectra employed throughout this work (see Methods).

Multilayer hBN has inversion (and 6-fold rotational) symmetry due to the natural 2H stacking of its van der Waals structure[21]. Any contribution at the second harmonic wavelength in few- to many- layer hBN is therefore restricted only to the broken inversion symmetry cases of interfaces and an odd number of layers and is inherently weak[17]. By conducting ultrafast pump-probe experiments we show that excitation of the IR-active TO ($E_{1u}$) phonon allows for the presence of FWM signals at energies of twice the probe photon plus or minus one phonon. Our simulations reveal the emergence of such an ultrafast, transient signal surrounding the second harmonic of an 800 nm probe pulse, as shown in Figure 3a. The signal on either side of harmonic order 2 highlights the shifting of the signal frequency up and down by the phonon energy in the two variations of the FWM process shown in Figure 3e. We note that the HHG spectra obtained in the presence of the TO excited phonon display additional signals along with the odd harmonics. This results mostly from the presence of phonon-induced sidebands, which are generated by electron and phonon frequencies (see discussion under Figure S1). The sideband effect also explains the dip at the even harmonic position in our simulations. The energy width between the two split peaks is approximately twice the energy of the TO mode, indicating that the nonlinearity is predominantly third-order.

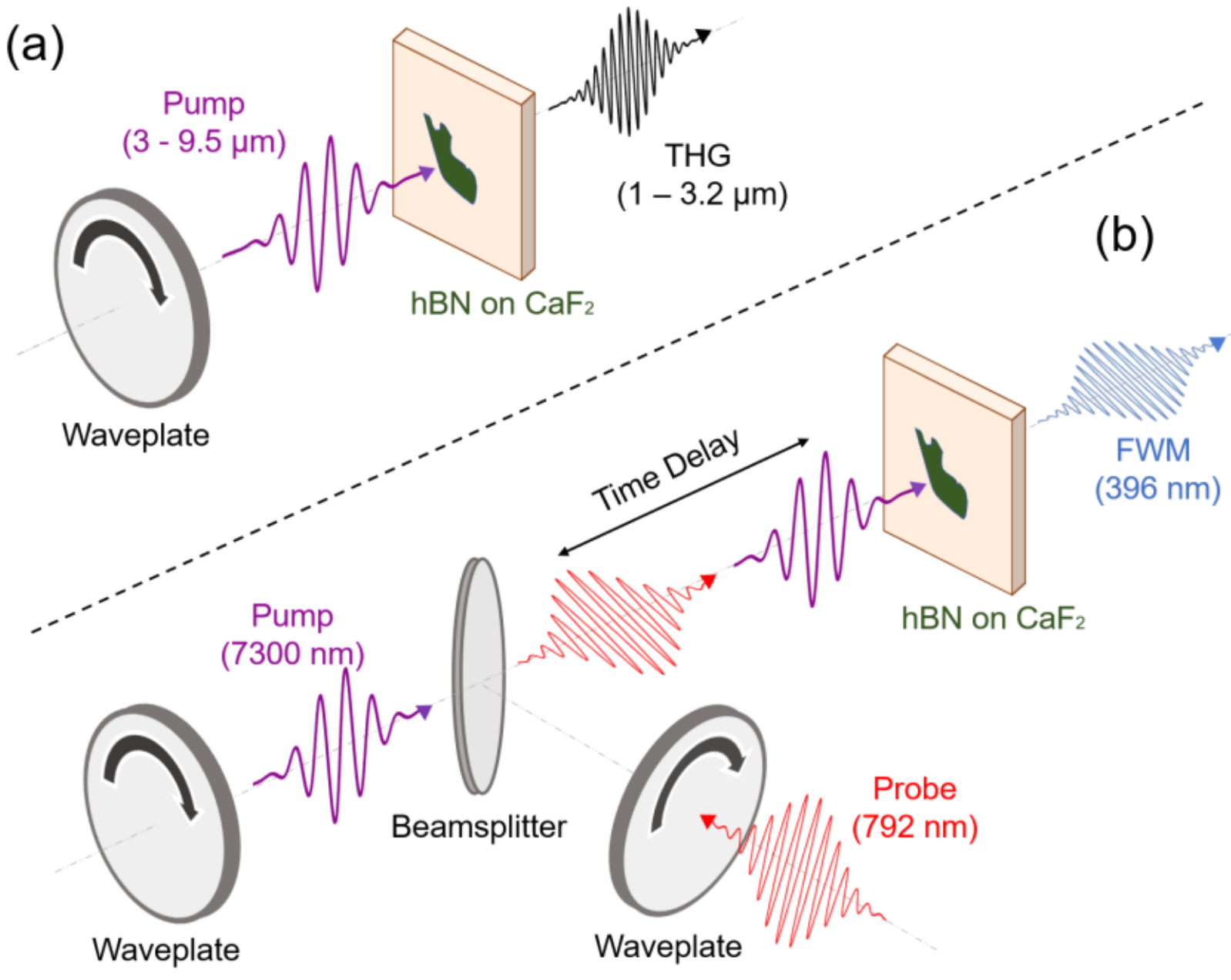

Figure 2: Setup for two experiments demonstrating phonon-enhanced nonlinearity in hBN, in transmission geometry. (a) Experimental setup for THG experiments. Detection is performed with PbS and MCT detectors, a lock-in amplifier, and boxcar-averaging. (b) Experimental setup for pump-probe FWM experiments. The time-delay is controlled by a mechanical delay stage with sub-1 μm step size. The pump and probe are both focused onto the sample with a reflective objective with 0.5 numerical aperture. Detection is performed with a silicon photomultiplier tube and lock-in amplifier.

In our experiments, the measured signal near the second harmonic wavelength of 396 nm is presented in Figure 3b as a function of the time delay between 792 nm and 7.3 μm pulses. The probe pulse from an amplified Titanium-Sapphire laser is scanned in time by a mechanical delay line relative to the pump pulse from a mid-infrared optical parametric amplifier and difference-frequency generation module. The powers and relative polarizations are set with filters and half-wave plates (HWP), and the two beams are then combined on a beamsplitter before being focused onto the sample by a reflective objective (the experimental setup is shown in Figure 2b, with further details in Methods). When the probe pulse precedes the pump pulse, no FWM signal is measured, indicating that the interface SHG and odd layer-number contributions are below the noise floor. The time-resolved signal displays a strong signal at the zero-time delay, when the probe pulse's arrival coincides with the excitation of the hBN phonon-polariton. The transient signal relaxes back to zero with a time constant of 120 fs, which is approximately twice the pump pulse duration. When pumped far off from the phonon resonance, no FWM signal is measured. Fast oscillations on the pump-probe trace provide a direct measurement of the oscillating atomic displacements in time. A pedestal on that same signal is a consequence of the finite response time of each peak being slower than the driving frequency. In Figures 3c and 3d we show the dependence of the FWM yield on the intensity of the probe and pump, respectively. A quadratic dependence of the FWM intensity on the probe power is observed, while a linear scaling of the signal with respect to the mid-IR intensity is found. Figure 3d reproduces the expected linear dependence on the mid-IR pump pulse based on the pair of $\chi^{(3)}$ interactions depicted in Figure 3e. We do not observe high-order phonon-resonant processes since the strength of such signals are below the sensitivity of the detection system.

We determine the dependence of the ultrafast FWM on the orientation of the pump and probe polarizations with respect to the crystal high-symmetry axes. Figure 4a gives the total normalized FWM yield for 360° rotation of both pulses (180° rotation is measured and the data is then mirrored). We observe a polarization behavior unique from either the inherent 6-fold $\chi^{(2)}$ or isotropic $\chi^{(3)}$ symmetries of purely electronic hBN nonlinearities[22]. Specifically, the emission appears to closely obey the functional form,

$$\mathrm{FWM}(\theta, \phi) = [\alpha \cos^2(3\theta) + \beta \sin^2(3\theta)] \cos^2(\theta - \phi) \quad , \qquad (1)$$

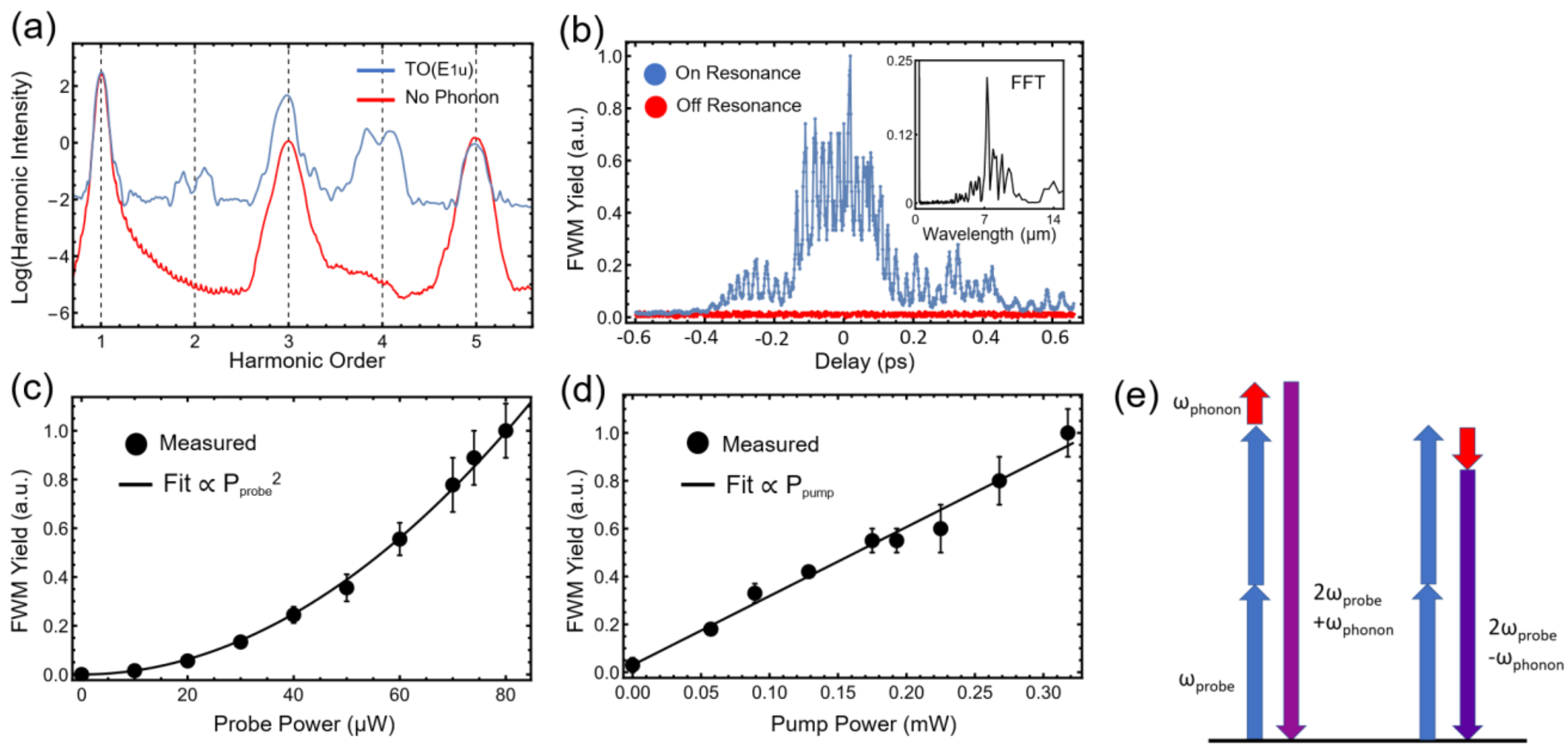

Figure 3: Four-wave mixing between a probe signal and the mid-IR pump at the phonon frequency. (a) TDDFT simulations show the emergence of FWM nonlinearity during resonant excitation of the TO phonon mode. (b) Time-resolved FWM yield (normalized) of the 792 nm probe pulse. While the pumps are temporally overlapped, an ultrafast third-order nonlinearity is measured. The transient signal vanishes following a 200-fs time constant, or about twice the pulse duration. The appearance of wings in the time-delay scan is a result of a non-perfectly Gaussian pulse, a result of strong atmospheric absorption. Inset: Fourier-Transform of the FWM time-delay. (c) Dependence of measured FWM yield on probe power. (d) Dependence of measured FWM on pump power. The FWM yield increases linearly with the phonon driving intensity. (e) Two versions of the proposed FWM process. The $2\omega_{phonon}$ energy difference in the emissions is consistent with the splitting observed in the theoretical spectra.

where θ and ϕ are the angles of the pump and probe relative to the zigzag (ZZ) axis of the crystal, respectively, and α and β determine the relative strengths of the emission along the ZZ and Armchair (AC) axes, respectively. The nonlinear yields peak only along ZZ axes that are being resonantly driven with a phonon-polariton. This is most clearly visible in the linecuts of the probe polarization dependence for pump fields aligned parallel to the AC and ZZ axes, given in Figure 4b. Even when the pump excitation is aligned with an AC axis, the two adjacent ZZ oriented $\mathrm{TO}(\mathrm{E}_{1u})$ phonons oscillate with a relatively small amplitude, and we observe phonon-mediated FWM, whereas the ZZ axis at exactly 90º from that excitation shows no emission. From Figure 4 we determine that the phonon-mediated FWM is at least 3 times greater parallel to ZZ than to the AC directions. This is supported by time-dependent density functional theory simulations in Figure 4c, which identifies new nonlinearity along both symmetry axes, though much greater for the $\mathrm{TO}(\mathrm{E}_{1u})$ phonon than the relative $\pi$ phase $\mathrm{LO}(\mathrm{E}_{1u})$.

**Phonon-enhanced third-harmonic generation:** When driven beyond the previously discussed weak-excitation regime, further enhanced nonlinearities emerge in hBN. We show the integrated and normalized experimental THG amplitudes for a range of pump wavelengths from 3 μm to 9.5 μm in Figure 5a as blue dots, which are in excellent agreement with the calculations discussed in Figure 1 and plotted as green dots in Figure 5a. The third-harmonic exhibits a strong peak for pump wavelengths near the TO phonon resonance at $\lambda$ = 7.3 μm, which is far from any electronic or excitonic resonances. We fit the data to a Lorentzian and extract a resonance full-width at half-maximum of 500 nm. THG yields are below the noise level for all $\lambda_{pump}$ less than 6 μm or greater than 9 μm, compared to that of the resonant signal which yields at least a 30-fold increase, and thus the phononic enhancement of the THG coefficient at the phonon-polariton wavelength is significantly greater than the purely-electronic component in this regime. In Figure 5b we plot the measured intensity dependence of the THG signal for $\lambda_{pump}$ = 7.3 μm. The fit to a cubic function indicates that the measured nonlinearity is third-order and that the scaling is perturbative, even at high intensities[23]. We note that a similar effect has been observed in the phononic second-harmonic generation of $\mathrm{LiNbO_3}$, which also remained in the perturbative regime at higher-than-expected intensities. Ultimately, significant enhancement of the phonon-induced nonlinearity could be further provided through use of subwavelength structures that support confined phonon-polaritons[6,24].

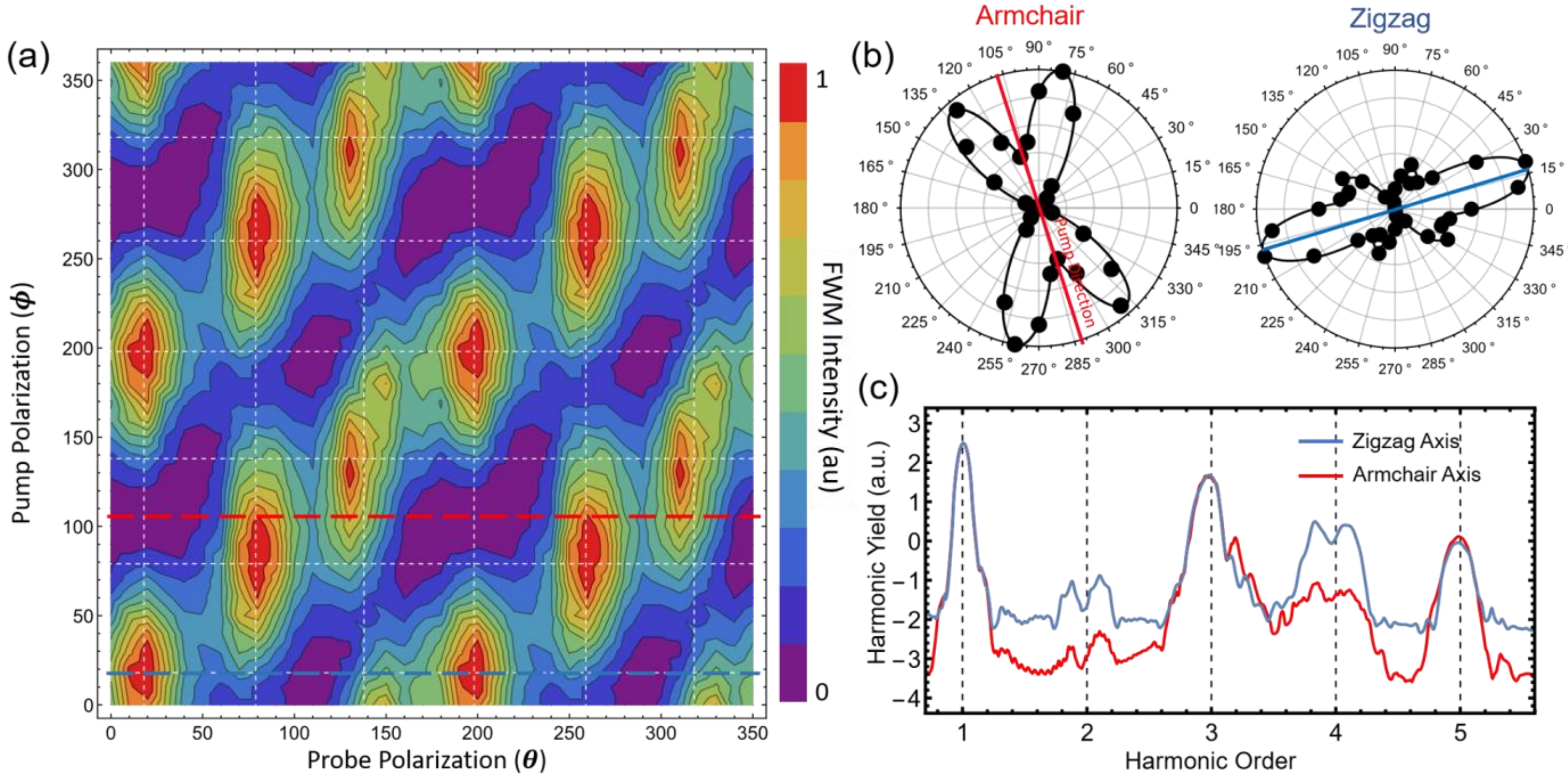

Figure 4: Polarization dependence of the FWM process. (a) Pump and probe polarization dependence of total FWM emission (normalized). White dashed lines indicate ZZ axes and are included as a guide for the eye and to emphasize the 60° periodicity. (b) Linecuts along the ZZ and AC axes from (a). Solid black lines are fit from Equation 1, with $\alpha$ and $\beta$ as fit parameters. (c) TDDFT-computed harmonic generation spectra for pump and probe pulses co-polarized along the ZZ (blue) and AC (red) directions. The TO ($E_{1u}$) phonon present along the ZZ axes leads to the greatest nonlinearity.

We also performed TDDFT simulations of the wavelength dependence of a higher-order harmonic (HHG) spectra of bulk hBN See Figure S1 for two different pump lasers with a wavelength of $\lambda_{pump}$ = 7.3 µm (polarized parallel to the TO mode) and $\lambda_{pump}$ = 6.2 µm (polarized parallel to the LO mode). Changing the wavelength and polarization of the pump laser can lead to the excitation of different phonon modes and lead to significant modulation of the HHG spectra. Excitation of either the TO or LO mode leads to noticeable modifications of the high-harmonic spectra, with the TO ($E_{1u}$) enhancement being one order of magnitude greater than that caused by LO excitation. Furthermore, more intense laser pulses introduce larger phonon amplitudes and lead to larger nonlinearity. As seen from a pump intensity of 2.5 x $10^{11}$ W/cm$^2$, the high-harmonic yields can be increased in a wide energy regime, and the high harmonic generation plateau is enhanced (Figure S2), which is attributed to the increased atomic movement and enhanced nonlinearity.

# Discussion

We have demonstrated greatly enhanced nonlinearities for optical parametric processes through resonant phonon driving. Furthermore, the appearance of fast oscillations in the pump-probe signal provides the capability for real-time monitoring of atomic motion and evolution driven by ultrafast laser pulses. The maximum achievable FWM efficiency is highly sensitive to the underlying symmetries of the hexagonal lattice, peaking along the ZZ axes where the greatest atomic displacements are known to occur. We extend the light-matter interactions confined by the hyperbolic nature of the hBN phonon dispersion to a strongly nonlinear regime by demonstrating that the large electron-phonon coupling leads to a nearly two order of magnitude enhancement of THG. We note that the phonon resonance present in this work is related to Floquet engineering[25,26]. Floquet engineering involves applying a periodic perturbation to a quantum system, creating a series of states that can be utilized to engineer various properties of the system. The effect of driving the phonon in the harmonic spectra can be interpreted as Floquet phonon engineering, where the harmonic oscillation of the phonon is the external driving frequency in the Floquet theory[26]. Efficient coupling of light to hBN phonon-polaritons at normal incidence places stringent requirements on the allowed optical excitation wavelength. For the free-space wavevector **k** = 0, the required photon wavelength of 7.3 μm is fixed, independent of flake thickness[13]. We only focus on the phonon mode at the Γ point because of energy and momentum conservation. Photon momentum is negligible compared to the size of the Brillouin zone of hBN. The lattice dynamics are driven by an external laser with a wavelength of 7.3 μm polarized parallel to the atomic displacements of the TO ($E_{1u}$) mode. Under those conditions, the simulated real-time evolution of the atomic displacements exhibits clear signatures of only the TO ($E_{1u}$) mode being excited, as shown in Figures 1b and S1a. Saturation of the THG yield below its perturbative cubic scaling was not observed and is more likely to occur closer to the onset of sample damage.

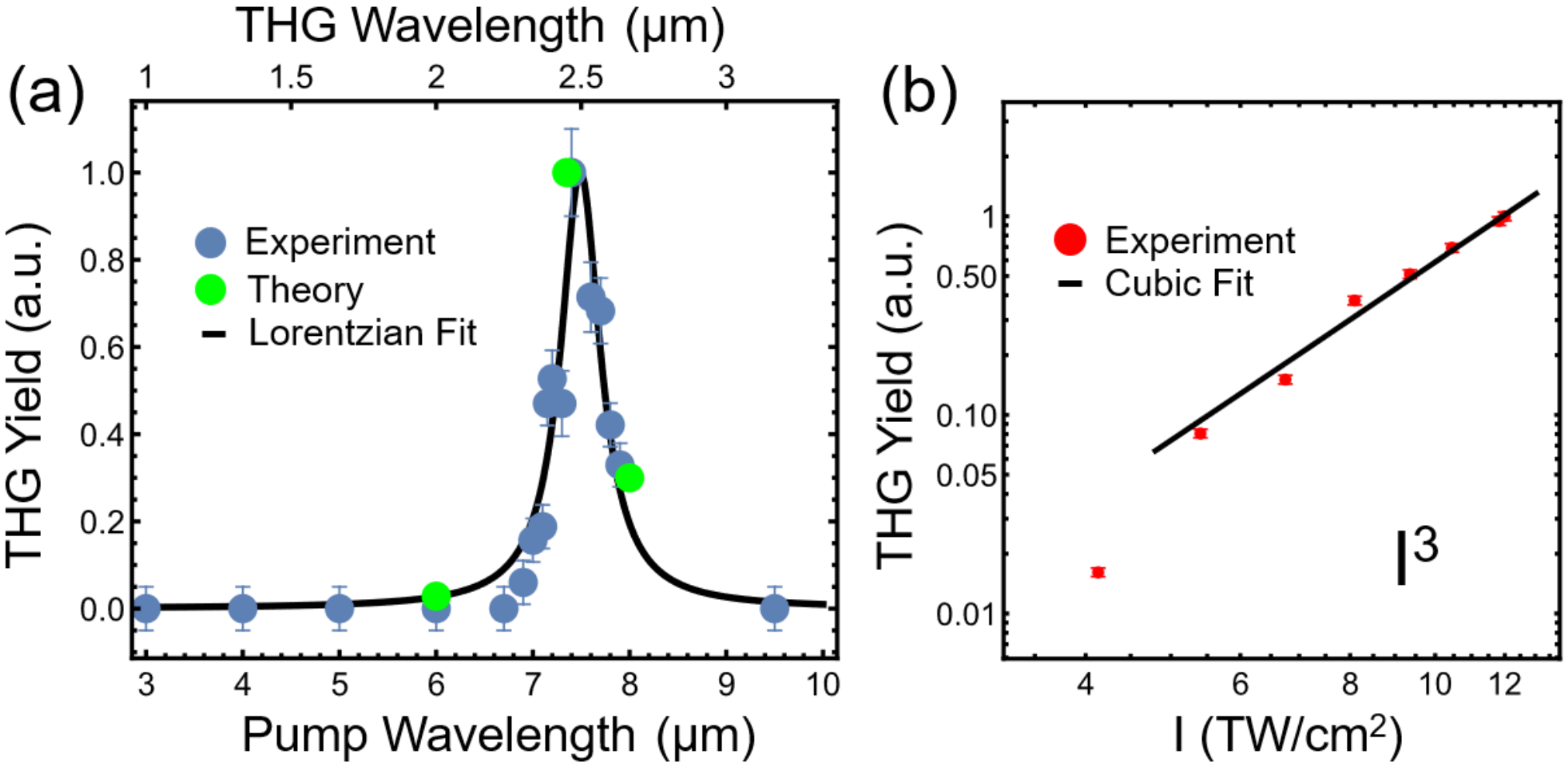

Figure 5: Wavelength and power dependence of the THG process. (a) Normalized third-harmonic generation yields of 120-fs pulses as a function of pump wavelengths throughout the mid-IR. THG yields are below the noise level for all wavelengths less than 6 μm or greater than 9 μm. Within a roughly 1 μm bandwidth of the phonon-polariton resonance, a THG enhancement of 30x is observed. The black line is a Lorentzian fit to the data (blue dots) with full-width at half-maximum of 500 nm. The green dots were obtained by integrating the third-harmonic signal in TDDFT simulations and show excellent agreement with experiments. (b) Normalized intensity dependence of THG pumped on-resonance at 7.3 μm and measured at 2.43 μm. The data is a close fit to $I^3$, indicating that the nonlinear process is scaling perturbatively.

# Methods

**Theory:** In the TDDFT simulations, the time-evolution of the wave functions and the evaluation of the time-dependent electronic current were computed by propagating the Kohn-Sham equations in real space and real time, as implemented in the Octopus code[27,28], in the adiabatic LDA[29] (the findings and trends discussed in the present work are robust with different functionals) and with semi-periodic boundary conditions. All calculations were performed using fully relativistic Hartwigsen, Goedecker, and Hutter (HGH) pseudopotentials[30]. The real-space cell was sampled with a grid spacing of 0.4 bohr and the Brillouin zone was sampled with a 42 × 42 × 21 k-point grid, which yielded highly converged results. The boron nitride bond length is taken here as the experimental value of 1.445 Å. The laser was treated in the dipole approximation using the velocity gauge (that implies that we impose the induced vector field to be time dependent but homogeneous in space), and we used a sin-square pulse envelope. In all of our calculations, we used a carrier-envelope phase of f = 0 [31]. The full harmonic spectrum is computed directly from the total electronic current **j**(**r**, t) as

$$HHG(\omega) = \left| FT\left(\frac{\partial}{\partial t}\int d^3\boldsymbol{r}\,\boldsymbol{j}(\boldsymbol{r},t)\right)\right|^2, \tag{2}$$

where FT denotes the Fourier transform. The atomic vibrations of phonon modes are prepared with the following two methods: (i) the time-evolution from a distorted atomic configuration along the phonon modes of 1% of the bulk hBN lattice. (ii) application of pump laser pulses with the same frequencies and polarizations of phonon modes. Our calculations confirm the two methods are equivalent in the simulations of high-harmonic generation.

**Experiments:** We performed the nonlinear experiments on high-quality hBN flakes with thicknesses of 10 to 50 nm and typical sizes of tens of microns. The flakes are exfoliated onto a $CaF_2$ substrate, chosen for its high transparency in both the visible and mid-infrared and its relatively small nonlinearity. For our long-wave infrared pump pulses we utilize an optical parametric amplifier (OPA, Light Conversion HE Topas Prime) pumped by an amplified Titanium-sapphire laser system (Coherent Legend Elite) operating at a 1-kHz repetition rate with 6 mJ of pulse energy. The OPA produces 60-fs duration signal and idler pulses with center wavelengths in the near-IR. The parametric amplifier output is then used to seed an additional difference frequency generation (DFG) module for all mid-infrared measurements from $\lambda_{pump}$ = 3 to 10 μm with pulse durations ranging from 70- to 120-fs. Pulse intensities are consistently set below the hBN damage threshold, which we estimate to be 50 TW/cm$^2$. For THG experiments the pump is focused onto an hBN flake using a 2-cm focal length $CaF_2$ lens, and the emitted THG signal is collected in a transmission geometry by an identical lens. After the residual pump beam is rejected by a short-pass filter, the remaining THG is measured on a PbS detector for harmonic wavelengths $\lambda_{THG}$ below 1.7 μm, and on a liquid nitrogen-cooled MCT detector for all $\lambda_{THG}$ greater than 2 μm.

For visible wavelength measurements we modify the experimental setup to a pump-probe scheme with the addition of a 792 nm, 45-fs pulse from the same amplified Ti-Sapphire laser. A variable time-delay separates the 7.3 μm pump pulse that we use to excite the phonon, from the near-IR probe which produces the FWM signal at 396 nm. The intensity of both pulses is maintained at or below the (TW/cm$^2$) range, which is below the hBN damage threshold. The time delay between pulses is controlled by a mechanical delay line with sub-1-μm step size. The polarization of the pump and probe beams are independently rotated with zero-order half-wave plates and wire-grid polarizers before the pulses are combined on a beamsplitter. The collinear pump and probe are focused onto an hBN flake using a reflective objective (NA = 0.5), which ensures the same focal plane for the two beams with very different wavelengths. The FWM signal produced by the 792 nm pulse can be collected either in the reflection geometry by the reflective objective, or in transmission by a $CaF_2$ lens. The signal is then directed through a bandpass filter with a 10-nm bandwidth to reject the residual 792 nm and 7.3 μm light before detection on a fast photomultiplier tube (PMT) and lock-in amplifier.

## Data Availability

The data generated during the study is available from the corresponding author upon reasonable request.

## Acknowledgements

This work is supported as part of Programmable Quantum Materials, an Energy Frontier Research Center funded by the U.S. Department of Energy (DOE), Office of Science, Basic Energy Sciences (BES), under award no. DE-SC0019443. The work of J.Z., LX., N.T., and A.R. was supported by the European Research Council (ERC-2015-AdG694097), the Cluster of Excellence ’CUI: Advanced Imaging of Matter’ of the Deutsche Forschungsgemeinschaft (DFG) - EXC 2056 - project ID 390715994, Grupos Consolidados (IT1249-19), partially by the Federal Ministry of Education and Research Grant RouTe-13N14839, the SFB925 ”Light induced dynamics and control of correlated quantum systems”, The Flatiron Institute is a division of the Simons Foundation. Support by the Max Planck Institute - New York City Center for Non-Equilibrium Quantum Phenomena is acknowledged. J.Z. acknowledges funding from the European Union’s Horizon 2020 research and innovation program under the Marie Sklodowska-Curie grant agreement No. 886291 (PeSD-NeSL). We thank I-Te Lu for the fruitful discussions. K.W. and T.T. acknowledge support from the Elemental Strategy Initiative conducted by the MEXT, Japan (Grant Number JPMXP0112101001) and JSPS KAKENHI (Grant Numbers 19H05790 and JP20H00354). C.Y.C. acknowledges support from the NSF Graduate Research Fellowship Program DGE 16-44869.

## Author Contributions

J.S.G., C.Y.C., M.J., and G.N.P. performed experiments. J.Z., N.T., and L.X. performed theory and simulations. Samples from K.W. and T.T. were prepared by S.C. Research was supervised by A.L.G., A.R., and J.H.

## Competing Interests

The authors declare no competing interests.